\documentclass[twocolumn,prl]{revtex4}

\usepackage{amsmath, amssymb, graphicx, comment, amsthm}
\usepackage{bbold, bm}

\newcommand{\be}{\begin{eqnarray}}
\newcommand{\ee}{\end{eqnarray}}

\newcommand{\bcf}{\begin{figure}}
\newcommand{\ecf}{\end{figure}}

\newcommand{\SU}{\text{SU}}

\newcommand{\su}{\mathfrak{su}}

\def\Tr{\text{Tr}}

\def\bea {\begin{eqnarray}}

\def\eea {\end{eqnarray}}

\def\A{{\bm A}}
\def\e{\bm{e}}
\def\X{\bm{X}}
\def\E{\bm{E}}
\def\F{\bm{F}}
\def\T{\bm{T}}

\def\t{\bm{\tau}}
\def\rd{\mathrm{d}}

\def\K{\bm{K}}
\def\L{\bm{\lambda}}
\def\XI{\bm{\xi}}
\def\bL{\bar{\bm{\lambda}}}
\def\bXI{\bar{\bm{\xi}}}
\def\bN{\bar{N}}

\def\y{\bm{y}}
\def\p{\bm{p}}
\def\q{\bm{q}}

\def\r{\mathsf{r}}

\def\n{\mathsf{n}}

\def\f{\mathsf{f}}

\def\b{\mathsf{b}}
\def\c{\mathsf{c}}

\begin{document}

\title{Discrete Hamiltonian for General Relativity}
\author{Jonathan Ziprick}
\email{jziprick@unb.ca}
\author{Jack Gegenberg}
\email{geg@unb.ca}
\affiliation{University of New Brunswick\\
Department of Mathematics and Statistics\\
Fredericton, NB E3B 5A3, Canada}
\date{\today}

\begin{abstract}
Beginning from the Ashtekar formulation of canonical general relativity, we derive a physical Hamiltonian written in terms of (classical) loop gravity variables. This is done by gauge-fixing the gravitational fields within a complex of three-dimensional cells such that curvature and torsion vanish within each cell. The resulting theory is holographic, with the bulk dynamics being captured completely by degrees of freedom living on cell boundaries. Quantization is readily obtainable by existing methods.
\end{abstract}

\maketitle

\section{Introduction}

The strong equivalence principle is one of the building blocks of general relativity (GR). It implies that an appropriate diffeomorphism can make spacetime flat at any point, while it remains generally curved elsewhere. This classical notion of locality however, is thought to extend only up to the Planck scale where various approaches to quantum gravity \cite{cdt, sorkin, LQG, spinfoam, qregge, string} suggest that a discrete spacetime structure emerges. The main challenge facing all of these programs is to develop a complete, self-consistent quantum dynamics that agrees with GR in the classical limit. This is the `bottom-up' view. Here we take the `top-down' view, beginning from the continuous theory to first derive a discrete, dynamical theory of GR {\it before} constructing the quantum theory \cite{LF}. The main idea is to extend local flatness to small but finite `cells' of space, over small (continuous) time intervals, and to impose this expicitly within the canonical formalism of GR.

We define spatial hypersurfaces in terms of a countable but arbitrary number of three-dimensional cells, constraining the geometrical fields defined within each cell to be intrinsically and extrinsically flat. This method yields an invertible map from the continuous to a discrete phase space written in terms of holonomy-flux variables for loop gravity \cite{FGZ}. These parameterize equivalence classes of continuous, piecewise-flat spatial hypersurfaces in terms of finite degrees of freedom (DOF), in a way which reproduces the full class of spatial hypersurfaces as the number of cells grows large \cite{T}. On the reduced phase space, the constraints within each cell are identically satisfied leaving a discrete theory written in terms of holonomies and fluxes on cell boundaries. Maintaining a continuous geometry dynamically completely fixes the gauge thereby providing a physical Hamiltonian. The result is a discrete theory of GR suitable for canonical quantization in terms of the operators and Hilbert space of loop quantum gravity (LQG).

\section{Conventions and Notation}
We use $\su(2)$ basis elements ${\bm \tau}^i$ (for $i=1,2,3$) which are given by $-i/2$ times the Pauli matrices. We define a trace notation such that $\Tr(\t^i \t^j) \equiv -2\mbox{trace}(\t^i \t^j) = \delta^{ij}$, and a bracket $[\cdot, \cdot]$ which implies taking the $\su(2)$ commutator \textit{and} wedge product between entries. Elements $\bm{v} \in \su(2)$ are written in bold font where $\bm{v} \equiv v^i {\bm \tau}^i$, and $\su(2)$ indices are all written `up' for convenience (since the Cartan-Killing metric is the Kronecker delta). Elements of $\su(2)$ represent vectors in $\mathbb{R}^3$ via $v^i \equiv \Tr (\t^i \bm{v})$, and we define the modulus $|\bm{v}| \equiv \sqrt{\Tr(\bm{v}\bm{v})}$. %in this context we use a dot product $\u \cdot \bm{v} \equiv \Tr (\u \bm{v})$ and a cross product $\u \times \bm{v} \equiv [\u, \bm{v}]$. Unit vectors are denoted with a hat $\hat{\bm{v}} \cdot \hat{\bm{v}} = |\hat{\bm{v}}|^2 = 1$.

\section{Continuous Hamiltonian}

A Hamiltonian for GR may be written in terms of the Ashtekar variables $(\A, \E)$ \cite{Ashtekar}. The momentum two-form, or `electric field', $\E$ describes the intrinsic geometry of spatial hypersurfaces, being given in terms of the frame-field $\e$ according to:
$
\E = \frac{1}{2} [\e, \e].
$
The configuration variable is the connection one-form $\A$. It contains both intrinsic and extrinsic information, being given by:
$
\A = \bm{\Gamma} + i \K,
$
where $\K$ is the extrinsic curvature and $\bm{\Gamma}$ is the (torsion-free) Levi-Civita connection.

At this point, the fields are defined on a simply connected three-manifold $\c$, which has a closed boundary $\partial \c$. The fields $(\A, \E)$ together with the manifold $\c$ define a spatial hypersurface of spacetime. We begin by analyzing $(\A,\E)$ within a single cell $\c$, before discussing how to glue multiple cells together later on.

The canonical action $I = \Omega - H$ is composed of the symplectic term $\Omega$ and the Hamiltonian $H$. The symplectic term is given by:
\be
\Omega = i \int_\c \ \Tr \ \dot{\A} \wedge \E ,
\ee
which implies that $(\A, \E)$ form a conjugate pair of phase space variables.
The Hamiltonian is a sum of the scalar, vector and Gauss constraints:
%\be
%\label{ham}
%H &=& S[N] + V[\XI] + G[\L] ,
%\ee
%which are given by:
\begin{subequations}
\label{continuous-H}
\be
\label{SV}
S[N] + V[\XI] &=& \Tr \int_\c \left(N + \XI \right) \e \wedge  \F,\\
\label{G}
G[\L] &=& \Tr \int_\c \L \rd_{\A} \E ,
\ee
\end{subequations}
where $N$, $\XI$ and $\L$ are Lagrange multipliers and
$
\F \equiv \rd \A + \frac{1}{2} [\A, \A]
$
is the curvature of the Ashtekar connection. The scalar constraint generates dynamics, the vector constraint generates diffeomorphisms and the Gauss constraint generates $\SU(2)$ gauge transformations. The fields $(\A, \E)$ and Hamiltonian (\ref{continuous-H}) define GR with vanishing cosmological constant.

In order that the variational principal be well-defined, we add the following boundary terms to the Hamiltonian:
\be
\label{SVBT}
S_B[N] + V_B[\XI] &=& \Tr \int_{\partial \c} \left( N + \XI \right) \e \wedge \A , \\
\label{GBT}
G_B[\L] &=& -\Tr \int_{\partial \c} \ \L \E .
\ee
and place boundary conditions on the electric field as follows. Points in $\c$ are labeled by a chart $x$, and for simplicity we choose this to be a set of cartesian coordinates in which $\c$ takes the form of a tetrahedron. This shape has no meaning -- the physically meaningful geometry is that described by $\E(x)$ which defines the spatial metric. We want this physical geometry to be a tetrahedron as well, but it is important to keep in mind that the tetrahedron defined by $\E(x)$ is generally different than the one described by the chart $x$. In order that $\E$ describes a tetrahedron, we impose the boundary condition that $\E$ is constant along each face. The value of this constant is left free, which leaves three DOF in $\E$ per face and allows the faces to move. This boundary dynamics is our main result and is developed in the remainder of the article.

\section{Boundary theory}

We now present a gauge reduction in which the canonical action vanishes within $\c$ and leaves a boundary theory on $\partial \c$. After studying a single three-manifold $\c$, we subsequently consider a collection of these cells which are glued together to form a larger, piecewise smooth three-manifold that is not simply connected. In the end we arrive at a physical Hamiltonian describing the full dynamics of such data, parameterized completely by DOF on the boundaries $\partial \c$.

\subsection{Kinematics}

We impose conditions such that $(\A, \E)$ describe an intrinsically and extrinsically flat geometry, as in the `flat cell gauge' of \cite{FGZ} which was further studied in \cite{FZ}. In terms of our phase space, this flatness implies \cite{FGZ} that the connection is flat $\F = 0$ and its torsion vanishes:
$
\T \equiv \rd \e + [\A, \e] = 0.
$
With these conditions the fields satisfy the constraints (\ref{SV}, \ref{G}) and therefore provide initial data for GR.

The solution to $\F = \T = 0$ is given in terms of an algebra-valued `coordinate' function $\y$ and a group-valued `rotation' function $a$:
\be
\label{solution}
\A = a^{-1} \rd a,  \qquad \qquad \e = a^{-1} \rd \y a .
\ee
The rotation function $a(x)$ is a holonomy defining parallel transport from some basepoint to any other point $x \in \c$ along an arbitrary path \footnote{The path is arbitrary because $\c$ is simply connected and the curvature vanishes everywhere $\F(\A)=0$ in $\c$.}. The coordinate function $\y$ defines the intrinsic geometry of $\c$.

In order to develop an invertible map between the continuous variables $(\A, \E)$ and the discrete boundary variables defined below, we choose a specific representative of the solutions (\ref{solution}) by defining $(a, \y)$ as follows. We choose each function $y^i(t,x)$ to be linear in $x$ so that the coordinate function provides a different, time dependant chart in which $\c$ appears as a tetrahedron. We also fix $\y = 0$ at the barycentre so that the tetrahedron is uniquely defined in terms of $\y$-coordinates. Notice that $\c$ may change shape according to $\dot{\y}$, while it is static in terms of the chart $x$.

To define the rotation function, we must first define some further structure. Consider a subdivision of $\c$ into four tetrahedron-shaped regions $\r_\f$, where each face $\f$ is the base of a sub-tetrahedron, and the barycentre $\n$ is at its apex \footnote{See Fig. 3 of \cite{Z} for an illustration of the lower-dimensional analog.}. Now consider a one-parameter family of two surfaces labeled by constant values of $\varphi(x)$ such that $\varphi = 0$ is the union of `lateral' faces (the faces which intersect at $\n$) and $\varphi = 1$ is the face $\f$. Notice however that $\varphi(x)$ is multiply-defined on edges (or `bones', in analogy with Regge geometry \cite{Regge}) of the tetrahedron.

In order to make $\varphi$ well-defined, we excise the bones as follows. Consider a cylinder of radius $\epsilon$ with a bone $\b$ along the axis, parameterized by angular and longitudinal coordinates $(\phi, z)$. The boundary is defined in the limit of $\epsilon \rightarrow 0$ by keeping the angular points distinct \footnote{This is the 3d analog of the smearing used in \cite{MW, Z}.}. This `smears' out the values of $\varphi$ over the boundary $\partial \b$ so that it is now well-defined, varying as one travels around $\partial \b$.

The bones are to be seen as topological defects, and must remain one-dimensional in terms of $\y$-coordinates. This requires the following boundary condition \cite{MW, Z}:
\be
\label{econ}
\left. \partial_\varphi \y \right|_{\partial \b} = 0 .
\ee

%In order to have a closed boundary $\partial \c$ around the space $\c$, we must include the portion of $\partial \b$ which intersect $\partial \c$. We define this as $(\partial \b)_\c \equiv \partial \b \cap \partial \c$ and use this to write the total boundary as the disjoint union:
%\be
%\partial \c  = \bigsqcup_\f \f \bigsqcup_\b (\partial \b)_\c.
%\ee

Now we are ready to define the rotation function. Using a bump function $f(\varphi)$ and a `twist' parameter $\p_\f(t)$ we define $a(x)$ for all $x$ in $\r_\f$ as:
\be
\label{represent}
a(t, \varphi) = \exp \left( {\p_\f(t) \int_0^\varphi \rd \tilde{\varphi} \ f(\tilde{\varphi})} \right),
\ee
for any point in a region $\r_\f$. Notice path-ordering is not required since $\p_\f$ is constant in $x$. Due to the properties of a bump function, one can see that the connection is smooth everywhere in $\c$ and vanishes on faces. Since the chart $x$ is time-independent, the entire dynamics of $a$ is contained within the twist parameters.

The excision defined above is key for two reasons: 1) It allows curvature to be supported on bones while the fields $(\A, \E)$ remain well-defined everywhere else; 2) treating these sources of curvature as topological defects prevents the curvature from spreading out as time evolves, so that the construction is preserved dynamically.

\subsection{Reduction}

The {\it discrete} boundary theory is obtained by using (\ref{solution}) as gauge conditions in the continuous theory. Maintaining these conditions throughout the evolution fixes the Lagrange multipliers, up to undetermined constants, in terms of $(a, \y)$ and $(\dot{a}, \dot{\y})$. The fully-determined, non-constant parts describe the local diffeomorphisms and rotations necessary to preserve the form of $(a, \y)$ given above. The undetermined components are written as $\bN$, $a^{-1} \bL a$ and $a^{-1} \bXI a$ where an overbar denotes a constant; these determine global transformations over $\c$ and will be fixed in the subsequent section.
%For such data the equations of motion are:
%\be
%i\dot{\A} = \rd_\A \L, \qquad i \dot{\E} = \e \wedge \rd N + \frac{1}{2}[\e, \rd_\A \XI] + [\E, \L].
%\ee
%Substituting $(\ref{solution})$ into the above provides equations which must be satisfied in order to preserve the gauge conditions throughout the evolution.

Let us now impose (\ref{solution}) within the theory by direct substitution in the canonical action. One finds that the terms (\ref{SV}, \ref{G}) each vanish identically, while the symplectic term vanishes in the bulk but leaves behind a boundary contribution \cite{FGZ, T}:
\be
\label{SP}
\Omega = - i \Tr \sum_\f h_\f^{-1} \dot{h}_\f \X_\f.
\ee
The four pairs $(h_f, \X_f)$ are a discrete set of boundary variables:
\be
\label{map}
h_\f \equiv a^{-1}(x), \qquad \qquad \X_\f \equiv \frac{1}{2} \int_\f [\rd \y, \rd \y] ,
\ee
for any point $x$ on $\f$. Each holonomy $h_\f$ defines parallel transport from a face $\f$ to the basepoint $\n$. Each `flux' $\X_\f$ defines a vector normal to $\f$ with a modulus equal to the area. The geometry of the tetrahedron $\c$ is determined by these four fluxes.

The discrete variables $\left(h_\f, \X_\f \right)$ parameterize a phase space \cite{Alekseev} on each face with the following Poisson brackets:
\be
&\left\{h_\f, h_{\f'} \right\} = 0, \qquad
\left\{X^i_\f, X^j_{\f'} \right\} = i \delta_{\f \f'} \epsilon^{ijk} X_\f^k,& \\
&\left\{h_{\f}, X^i_{\f'} \right\} = i \delta_{\f \f'} \t^i h_\f ,&
\ee
where $\f$ and $\f'$ are any two faces on the boundary $\partial \c$.

Let us now turn to the boundary terms. The fully-determined parts of the multipliers lead to terms which do not contribute to the discrete theory. By direct calculation, the undertermined components yield for (\ref{GBT}):
\be
\label{closure}
G_B[\bL] = - \Tr \bL \sum_\f \X_\f .
\ee
This is the well-known discrete Gauss (or closure) constraint which tells us that the fluxes add up to zero, as required for a closed two-surface.

Looking at the boundary terms (\ref{SVBT}), we find that these vanish on faces because $\A$ is constant along each face. On the two-surface $\partial \b$ surrounding each bone however, we find:
\be
\label{SVBT2}
S_B[\bN] + V_B[\bXI] = \Tr (N + \bXI) \sum_\f \sum_{\b: \partial \b \cap \f \ne 0} \q_\b \p_\f.
\ee
The first sum is over faces, and the second is over bones which intersect a face.
This simple form is obtained using that:
\be
\int_{\partial \b \cap \r_\f} \rd \varphi \ \partial_\varphi a a^{-1} = \p_\f,
\ee
and defining a vector:
\be
\q_\b := \int_{\partial \b \cap \r_\f} \rd z \ \partial_z \y ,
\ee
which gives the length and orientation of the bone.

In order to make use of (\ref{SVBT2}) in the boundary theory, we need to write it in terms of the phase space variables $(h_\f, \X_\f)$. This is possible because of the way in which the fields $(a, \y)$ have been defined. The $\q_\b$ define edges of a tetrahedron, and these can be written explicitly in terms of fluxes $\X_\f$ using simple geometrical arguments. For the twists, note that an element of the group $\SU(2)$ is the exponential of an $\su(2)$ algebra element $h_\f = e^{\p_\f}$, where $\p_\f$ defines an angle $|\p_\f|$ and axis of rotation. This map is invertible for $0 \le |\p_\f| < 2 \pi$.

The canonical theory for $(\A, \E)$ in $\c$ has been expressed entirely in terms of four sets of boundary variables $(h_\f, \X_\f)$. This discrete theory has the following Hamiltonian:
\be
\label{HB}
H_\c = S_B[\bN] + V_B[\bXI] + G_B[\bL],
\ee
where each term is given above in (\ref{closure}, \ref{SVBT2}); the vectors $\q_\b$ are explicit functions of fluxes, and the twists $\p_\f$ are explicit functions of the holonomies. At any given time, knowing the holonomies and fluxes allows one to unambiguously determine fields $(\A, \E)$ from (\ref{solution}), using that $y^i$ are cartesian coordinates for $\c$ with $y^i =0$ at the barycentre, and $a$ is given by (\ref{represent}). Therefore, evolution in the bulk is completely determined by (\ref{HB}) in this reduced theory.

For a single manifold $\c$, the \textit{holographic} correspondence described above holds for any choice of multipliers $(\bN, \bXI, \bL)$ which are spatially constant in $\c$. In the next section we glue many $\c$'s together to form a larger space. Requiring that the resulting geometry is continuous at all times fixes the value of each multiplier, within each $\c$.

In the boundary theory, variables $(h_\f, X_\f)$ are associated to each of the four faces in $\partial \c$ giving $6 \times 4 = 24$ (complex-valued) DOF. These are subject to the (first class) closure constraint which removes six DOF and leaves $18$ phase space DOF to describe the evolution. It is interesting that this is the same number of DOF that the Ashtekar variables possess at each point in the continuous theory, before taking gauge freedom into account. This suggests an analogy between points in the continuous theory and cells in the discrete theory.

\section{Gluing}
We now generalize to the case where $\c$ is but one of many `cells' in a {\it CW complex} \cite{Hatcher}, each constructed in the manner described above. In order that the fields $(\A, \E)$ describe a continuous geometry throughout the complex, certain gluing conditions must hold so that the triangles shared by adjacent tetrahedra match up. These are gauge conditions, and as we will see, maintaining these conditions dynamically fixes a unique solution for $(N, \bXI,  \bL)$. We refer the reader to \cite{Z} for a more detailed description (with illustrations) of a similar gluing in two dimensions, with the caveat that there are important differences due to the trivial nature of three-dimensional gravity.

To keep the presentation simple, let us consider a CW complex with only two cells $\c$ and $\c'$, where each face $\f \in \partial \c$ is identified with a face $\f^\prime \in \partial \c'$. The results presented below generalize to an arbitrary number of cells.

Within each cell the fields satisfy $\F = \T = 0$, and the solution is given in terms of local functions $(a_\c, \y_\c)$ and $(a_{\c'}, \y_{\c'})$. In order that the geometry is continuous, both cells must agree on the value of $\E$ at each shared face. From the form of $\E$ given by (\ref{solution}), this implies that differentials of the coordinate functions must agree, up to a rotation, as one approaches each face from either side.
%\be
%\label{gluing}
%\lim_{x \to \f} \ h_{\f} \ \rd \y_{\c}(x) \ h_{\f}^{-1} = \lim_{x' \to \f'} \ h_{\f'} \ \rd {\y}_{\c'}(x') \ h_{\f'}^{-1},
%\ee
%where the point $x \in \f$ is identified with $x' \in \f'$, and we used that $a_c(x) \to h^{-1}_\f$ as $x \to \f$.
This provides four conditions, one for each face. Note that $\A$ goes to zero smoothly as one approaches a face from either side, so that no gluing is required for a continuous connection.

These conditions imply that the flux $\X_{\f}$ as seen from $\c$ is related to the flux $\X_{\f'}$ seen from $\c'$ by:
\be
\label{fluxglue}
\X_{\f} =  - g_{\f} \X_{\f'} g_{\f}^{-1}.
\ee
This `gluing' condition fixes the areas of $\f$ and $\f'$ to be equal, i.e. $|\X_\f| = |\X_{\f'}|$. The relative orientation is a free variable $g_\f = h_{\f}^{-1} h_{\f'}$ describing parallel transport from $\c$ to $\c'$ through the face $\f \equiv \f'$.% It is important to note that (\ref{fluxglue}) combines $(h_\f, \X_\f)$ and $(h_{\f'}, \X_{\f'})$ into a single phase space parameterized by $(\X_\f, g_\f)$, with the same Poisson brackets \cite{FGZ, T}.

Although the areas of each identified pair $(\f, \f')$ are now set equal, the \textit{shapes} are generally different, so at this point we have a (discontinuous) `twisted' geometry \cite{twisted}. To obtain a continuous geometry we must also impose shape-matching conditions \cite{shape} which identify angles shared by $\f$ and $\f'$ so that the triangles represented by each face are the same. With these conditions and (\ref{fluxglue}), the fluxes now define a spatial Regge geometry.

Let us count the number of shape matching conditions needed on this two-cell CW complex. A single tetrahedron is defined by six parameters, so six conditions in total are needed to give each tetrahedron the same shape. We have given four above in (\ref{fluxglue}) to match the areas, so we require \textit{two} shape-matching conditions to obtain a continuous geometry across each face in the two-cell CW complex.

With the gluing and shape-matching conditions we have obtained a continuous geometry everywhere on the CW complex. Each tetrahedron-shaped cell is parameterized by $(h_\f, \X_\f)$ subject to these conditions and the closure constraint. The dynamics is generated by a Hamiltonian (\ref{HB}) local to each cell, which can be written explicitly in terms of the holonomy-flux variables defining each cell.

The number of DOF remaining after gluing are counted as follows. Consider two (so far unglued) cells for a total of eight faces, with a six parameter phase space $(h_\f, \X_\f)$ on each face. We must then subtract: two closure constraints (one for each cell); four gluing conditions (one for each pair of faces); two shape-matching conditions (one for each cell). The closure constraints and gluing conditions each contain three equations while shape-matching is a single equation; each equation removes a pair of phase space DOF. Considering all of this, the total DOF are $6\times (8 - 2 - 4) - 2 \times 2 = 8$. For an arbitrary number of cells one finds four DOF per cell, the same number of physical DOF as GR has per point. 

The previous paragraph implies that gluing cells together to form a continuous geometry completely reduces the theory. The covariantly-constant parts $(N_\c, \bXI_\c, \bL_\c)$ of the Lagrange multipliers must then have been fixed by the gluing and shape-matching conditions. The equations which do so come from the time derivatives of these conditions, using the Hamiltonian equations of motion for $\dot{\X}_\f$ and $\dot{h}_\f$. In the two-cell complex these total $14$ equations, linear in $(N_\c, \bXI_\c, \bL_\c)$, which exactly fix these $14$ multipliers. For an arbitrary number of cells, one obtains seven equations per cell to fix seven multipliers per cell. Note that these solutions are highly non-local, involving data from every cell in the complex. Notice also that after gluing, the analogy still holds between points in the continuous theory and cells in the discrete theory.

\section{Discussion}
We have obtained a discrete Hamiltonian for gravity from a reduction of the continuous theory. Beginning from the Ashtekar variables $(\A, \E)$ for canonical GR, we reduced to fields which are flat and torsion free in a piecewise manner, satisfying $\F=\T=0$ within each cell of an CW complex. For such a geometry, curvature and torsion have support only upon the one-skeleton, i.e. the bones. Gluing the cells together to form a continuous spatial geometry, and maintaining continuity at all times, gives conditions which completely fix the gauge. This yields a physical Hamiltonian (\ref{HB}) which generates dynamics in the discrete phase space. The resulting theory is not an approximation, but rather an exact description of dynamical, piecewise-flat and piecewise-torsion-free geometries.

%In the limit of many cells, such geometries describe the full class of continuous geometries. The holographic nature of gravity is exhibited in this reduction, with the bulk theory being described completely by boundary variables. The DOF counting suggests an analogy between points in the continuous theory and cells in the discrete theory. 

The discrete theory is written within the holonomy-flux phase space of loop gravity, from which the Hilbert space and operators of LQG \cite{LQG} can be constructed. These well-established techniques can be readily applied here --- the new feature is a Hamiltonian operator coming from the quantization of (\ref{HB}). To help develop this operator, the discrete theory serves as a valuable consitency check since the quantum dynamics must reproduce the classical dynamics in the $\hbar \to 0$ limit. Considering then that the discrete phase space in the continuum limit (when the number of cells grows large) describes the usual spatial geometries for canonical GR \cite{T}, quantization of the discrete theory is a promising avenue toward quantum gravity.

\begin{acknowledgments}
The authors wish to thank Viqar Husain and Gabor Kunstatter for helpful discussions and for useful comments on an earlier draft. J.Z. is especially grateful to his Ph.D. supervisor, Laurent Freidel, for teaching this perspective of LQG which emphasizes the interpretation of spin networks as spatial geometries. This work was supported by NSERC of Canada and an AARMS Postdoctoral Fellowship to J.Z.
\end{acknowledgments}

\end{document}